\documentclass[a4paper,11pt]{article}
\usepackage{jcappub}
\usepackage[T1]{fontenc}
\usepackage{amsmath, amssymb, graphics}
\usepackage{mathtools}
\usepackage{tablefootnote}
\usepackage{soul}
\usepackage{ulem}
\usepackage{color}
\usepackage{float}
\usepackage{makecell}
\usepackage{booktabs}
\usepackage[usenames,dvipsnames,svgnames,table]{xcolor}
\usepackage{indentfirst}
\usepackage{leftidx}
\usepackage{multirow}
\usepackage{comment}

\definecolor{LightCyan}{rgb}{0.88,1,1}
\newcolumntype{a}{>{\columncolor{LightCyan}}c}

\newcommand{\apj}{Astrophys. J.}

\newcommand{\jcap}{JCAP}
\newcommand{\physrep}{Phys. Reports}

\usepackage{xspace}

\title{Effects of the Cosmic Neutrino Background Capture on Astrophysical Objects}

\author[1]{Beatriz Hern\'andez-Molinero,}
\author[2,3]{Raul Jimenez,}
\author[1,4]{Carlos Pe\~na Garay,}

\affiliation[1]{Laboratorio Subterr\'aneo de Canfranc, 22880 - Estaci\'on de Canfranc, Huesca, Spain.}
\affiliation[2]{ICCUB, University of Barcelona, Marti  i Franques 1, E-08028 Barcelona, Spain.}
\affiliation[3]{Instituci\`o Catalana de Recerca i Estudis Avan\c{c}ats, Pg. Lluis Companys 23, Barcelona, E-08010, Spain.}
\affiliation[4]{I2SysBio, CSIC-University of Valencia, 46071 - Valencia, Spain.}

\emailAdd{bhernandez@lsc-canfranc.es; raul.jimenez@icc.ub.edu; cpenya@lsc-canfranc.es}

\abstract{Low-energy neutrinos from the cosmic background are captured by objects in the sky that contain material susceptible of single beta decay. Neutrons, which compose most of a neutron star, capture low-energy neutrinos from the cosmic neutrino background and release a high-energy electron in the MeV range. Also, planets contain unstable isotopes that capture the cosmic neutrinos. We show that this process is feasible and results in a non-negligible flux of electrons in the MeV range in neutron stars. We present a novel observable, the redshift evolution of the temperature of neutron stars due to neutrino capture, that could provide a route for detection of the cosmic neutrino background from future gravitational waves observatories. For planets the flux is significantly smaller and a measurement is not possible with currently envisioned technology. While the signature from neutron stars is small and challenging, it could result in a novel way to detect the cosmic neutrino background.}

\begin{document}

\maketitle

\section{Introduction}

In the standard $\Lambda$ Cold Dark Matter ($\Lambda$CDM) model of cosmology, neutrinos decouple from the primordial plasma approximately one second after the Big Bang, giving rise to the Cosmic Neutrino Background (C$_{\nu}$B) \cite{Neutrino-cosmology}. Despite being a robust prediction of both particle physics and cosmology, the C$_{\nu}$B has yet to be (directly) detected; this is due to the weakly interacting nature of neutrinos, which makes it an extremely elusive background. Detecting and characterizing the C$_{\nu}$B would provide a unique probe of the Universe at an epoch inaccessible to the Cosmic Microwave Background (CMB), offering insights into its conditions just one second after the Big Bang. Additionally, the C$_{\nu}$B detection could help determine whether neutrinos are their own antiparticle (\cite{Roulet:2018fyh,Hernandez-Molinero:2022zoo,2024JCAP...01..006H}, and references therein). If neutrinos are Majorana particles, both neutrinos and antineutrinos would contribute to the capture rate, nearly doubling the expected detection signal compared to the Dirac case, where only neutrinos would be captured. Understanding the nature of neutrinos is crucial for addressing fundamental questions such as the origin of the baryon asymmetry in the Universe. The most direct approach to probing the generation of lepton asymmetry  \cite{Fukugita:1986hr} is through neutrinoless double-beta decay experiments \cite{Goeppert-Mayer,furry}. However, if the total neutrino mass is as small as current cosmological constraints suggest ($ < 0.1$ eV) \cite{Planck:2018vyg}, these experiments would require detector masses on the order of 1–10 tons of the relevant isotope, posing significant experimental challenges. This has motivated the exploration of alternative detection methods, including the detection of the C$_{\nu}$B. 

Current efforts to detect the C$_{\nu}$B focus on ground-based tritium capture experiments \cite{Long_2014,Ptolemy}, which are still in the development phase. In this work, we explore an alternative approach: the potential detection of the C$_{\nu}$B through its capture by astronomical objects like neutron stars and planets. These will emit electrons in the MeV range, which could be observed by space-based telescopes. This mechanism could provide a novel and complementary pathway to probe the cosmic neutrino background and its fundamental properties. The paper is structured as follows: first, we describe the neutrino capture process in astrophysical objects. Then we apply it to the case of neutron stars in \S~\ref{sec:neutron}. We then study the case of capture by planets in \S~\ref{sec:planets}. We end with our conclusions and discussion.

\section{Cosmic neutrino capture process}

Cosmic neutrino momentum distribution is frozen since the first second after the Big Bang, when neutrinos decoupled from matter and freely streamed out. Nowadays, the temperature of the neutrino momentum Fermi-Dirac distribution is $T_{C_\nu B} = 1.95\, \rm{K}$ \cite{RouletBook,lesgourgues_pastor_2006}. This temperature implies a mean neutrino momentum of $\langle p_\nu\rangle=0.53\, \rm{meV}$. This very small momentum, along with the fact that the neutrino mass is smaller than $0.1 \,\rm{eV}$, implies that the cosmic neutrino capture can only take place in nuclei with no energy threshold in the capture reaction. Beta decaying nuclei are the good candidates because there is no energy threshold to capture the neutrino \cite{A.G.Cocco_2008}. The simplest case is the neutron-neutrino capture process:
\begin{equation}
    \nu_j + n \rightarrow p + e^- .
    \label{eq:capture process}
\end{equation}

Since our study explores observables sensitive to the C$_\nu$B by astrophysical objects, the incoming neutrino is represented in its mass eigenstate because cosmic neutrinos, having decoupled approximately one second after the Big Bang, propagate freely in their mass eigenstates.  

The differential cross-section for the capture process \eqref{eq:capture process} is given by:
\begin{equation}
\begin{split}
\frac{d\sigma_j (s_\nu,\,q_\nu)}{d\cos\theta}=& \frac{G_F^2 |V_{ud}|^2 |U_{ej}|^2}{4\pi}\frac{m_p}{m_n}\frac{p_e\, E_e}{v_{\nu_j}} F(Z,E_e)\times\\
&\left[\left(f^2 + 3g^2\right)A(s_\nu) + \left(f^2 - g^2\right)B(s_\nu) v_e \cos\theta\right]
\end{split}
\label{eq:differential cross section}
\end{equation}
where $f = 1$ and $g = 1.2695$ \cite{PhysRevD.110.030001}, corresponding to the case of a free neutron target. 

For the present analysis, the following neutrino mass eigenvalues are chosen:
\begin{equation}
    \begin{aligned}
    m_1 &= 1 \rm{meV} \\
    m_2 &= 8 \rm{meV} \\
    m_3 &= 50 \rm{meV} .
\end{aligned}
\label{eq:nu masses}
\end{equation}

Assuming a Fermi-Dirac distribution of momentum fixed by the C$_\nu$B temperature, $T_{\rm{C}\nu\rm{B}}=1.95\,\rm{K}$, the total cross-sections are computed as $\sigma_\nu^{\rm{left}}=9.3\cdot10^{-43}\rm{cm}^2$ for left-handed neutrinos and $\sigma_\nu^{\rm{right}}=7.9\cdot10^{-43}\rm{cm}^2$ for right-handed neutrinos. The detailed breakdown of these values is presented in Table \ref{tab:cross-sections}, along with the quantity $\sigma_\nu\cdot v_\nu/c$, which is relevant for rate calculations. The computation does not include averaging over neutrino spin to explicitly preserve the dependence on incident neutrino helicity. However, non-polarization is assumed for the electron and neutron.

\begin{table}[H]
    \centering
    \begin{tabular}{c|cc|cc}
    \hline
        \multirow{2}{*}{mass} & \multicolumn{2}{c|}{left} &  \multicolumn{2}{c}{right}\\
        & $\sigma_{\nu_j}$ $\left[\rm{cm}^2\right]$ & $\sigma_{\nu_j}\cdot v_{\nu_j}/c$ $\left[\rm{cm}^2\right]$ &   $\sigma_{\nu_j}$ $\left[\rm{cm}^2\right]$ & $\sigma_{\nu_j}\cdot v_{\nu_j}/c$ $\left[\rm{cm}^2\right]$ \\
    \hline
        $1\,\rm{meV}$ & $1.97\cdot10^{-43}$ & $7.17\cdot10^{-44}$ & $9.76\cdot10^{-44}$ & $2.82\cdot10^{-44}$\\
        $8\,\rm{meV}$ & $5.10\cdot10^{-43}$ & $2.39\cdot10^{-44}$ & $4.65\cdot10^{-43}$ & $2.09\cdot10^{-44}$\\
        $50\,\rm{meV}$ & $2.27\cdot10^{-43}$ & $1.68\cdot10^{-45}$ & $2.24\cdot10^{-43}$ & $1.64\cdot10^{-45}$\\
    \hline
        TOTAL & $9.35\cdot10^{-43}$ & $9.73\cdot10^{-44}$ & $7.87\cdot10^{-43}$ & $5.08\cdot10^{-44}$\\
    \hline
    \end{tabular}
    \caption{Computed cross-sections and cross-section times velocity for neutrino-neutron capture processes \eqref{eq:capture process} for each chosen neutrino mass eigenstate \eqref{eq:nu masses}.}
    \label{tab:cross-sections}
\end{table}

Once the cross-section has been calculated, the capture rate is computed in the following way:
\begin{equation}
\Gamma_{C\nu B}=\sum_j\left[\sigma_j(s_\nu=-1/2)v_{\nu_j}n_j(\nu_{hL}) + \sigma_j(s_\nu=+1/2)v_{\nu_j}n_j(\nu_{hR})\right]N_T,
\end{equation}
where $n_j(\nu_{hL})$ and $n_j(\nu_{hR})$ are the densities for left- and right-handed cosmic neutrinos, respectively. If neutrinos are Majorana particles, the capture rate will be calculated by summing both left-handed and right-handed contributions, whereas only the left-handed contribution will be considered if neutrinos are Dirac particles. For simplicity, we assume
\begin{equation}
    \begin{aligned}
        & n_j(\nu_{hL}) =  56 \,\rm{cm}^{-3}, \ \ n_j(\nu_{hR}) = 56 \,\rm{cm}^{-3} \ \ \textrm{for Majorana case}\\
        & n_j(\nu_{hL}) = 56 \,\rm{cm}^{-3}, \ \ n_j(\nu_{hR}) = 0 \,\rm{cm}^{-3}\ \ \ \, \textrm{for Dirac case.}
    \end{aligned}
\end{equation}
However, it is worth noting that previous works \cite{Hernandez-Molinero_2024,Hernández-Molinero_2024_halos} have shown that these densities can be altered by gravitational effects from the detector's surroundings; nevertheless, in this paper, we restrict our analysis to the standard case. Lastly, $N_T$ represents the total number of nuclei within the target. The aim of this study is to maximize the capture rate by identifying astrophysical objects whose constituents can enhance the number of target nuclei, thereby increasing the number of neutrino captures.

\section{Neutron Stars}
\label{sec:neutron}

The first proposed astrophysical object is a neutron star where free neutrons in a very dense medium can be found.

From the cross-section, the mean free path of neutrinos inside a neutron star is determined using:
\begin{equation}
    \lambda_\nu=\frac{1}{\sigma_\nu\,\rho_{\rm{NS}}}
\end{equation}
where $\rho_{\rm{NS}}$ is the neutron star density. Approximating the neutron star density as the nuclear density $\rho_0 = 0.16\, \rm{fm}^{-3}$, the mean free path of a cosmic neutrino inside a neutron star is found to be $\lambda\sim100\,\rm{m}$. This implies that neutrino capture occurs in the outer layers of the neutron star (neutron stars are characterized by typical masses around $1.4\,M_\odot$ with radii of only about 10 km \cite{NS-estructure}). For further analysis, we focus on the inner crust, where neutron drip occurs and the concentration of free neutrons is significantly larger \cite{Carroll_Ostlie_2017}. 

Electrons produced via neutrino capture are expected to lose energy rapidly within the neutron star medium. The mean free path of an electron in a neutron-rich medium at nuclear density is calculated using the electromagnetic cross-section:

\begin{equation}
    \frac{\sigma^{EM}}{d\Omega} = \frac{\alpha^2\cos^2\left(\theta/2\right)}{4E^2\sin^4\left(\theta/2\right)}\frac{E'}{E}\left[\frac{{G^E_n}^2+\tau{G^M_n}^2}{1+\tau}+2\tau{G^M_n}^2\tan^2\left(\theta/2\right)\right]
\end{equation}
where $Q^2 = 4EE'\sin^2\left(\theta/2\right)$ and $\tau=\frac{Q^2}{4M_n^2}$. We have assumed a dipole form for the neutron form factor
\begin{subequations}
\begin{alignat}{2}
    G^E_n(Q^2) &= 0\\
    G^M_n(Q^2) &= \frac{\mu_n/\mu_N}{\left(1+\frac{Q^2}{M_V^2}\right)^2}
\end{alignat}
\end{subequations}
where $M_V^2 = 0.73 \,\rm{GeV}^2$. For the calculation, we have also assumed monoenergetic $1.2\,\rm{MeV}$ outgoing electrons because scattering angle and neutrino mass dependencies are negligible $\left(\Delta\sigma_{p_e}(\theta)=0.15\,\rm{eV}\right.$, $\left. \Delta\sigma_{p_e}(m_j)=0.02\,\rm{eV} \right)$. Therefore, the resulting cross-section is $\sigma^{EM}\sim10^{-30}\rm{cm}^2$, corresponding to a mean free path of $\lambda_e\sim10^{-11}$m. Even the ocean and atmosphere, which have thickness of approximately 3 cm and 1000 cm, respectively, and densities comparable to standard matter density ($10^6$ and $1\, \rm{g}/\rm{cm}^3$, respectively) will trap the electrons, as they are stopped within a few centimeters. Thus, electrons cannot escape the neutron star but instead deposit their energy within its interior.

\subsection{Redshift Evolution of Neutron Star Temperature}
Therefore, given that electrons are trapped inside the neutron star, we will quantify the energy deposition within an outer shell of 2 km thickness. Within this volume we can ensure that nuclear density is reached \cite{Carroll_Ostlie_2017,PhysRevD.99.124029} and hence cosmic neutrinos will not go further than this depth.  

The capture rate in this volume is given by:
\begin{equation}
    \Gamma=\sum_j{\frac{4\pi}{\left(2\pi\right)^3}\int_0^\infty {dq_{\nu_j}\frac{q_{\nu_j}^2}{1+\exp(q_{\nu_j}/T_{\rm{C}\nu\rm{B}})}}
    \int_0^{4\pi}\frac{d\sigma_{\nu_j}}{d\Omega_e}\,d\Omega_e\, v_{\nu_j}\,N_n}
    \label{eq:rate}
\end{equation}
where $v_\nu$ is the neutrino velocity described by the Fermi-Dirac velocity distributions with $\rm{mass} = m_j$ and $T=T_{\rm{C}\nu\rm{B}}$ and $N_n = \rho_{\rm{NS}}\cdot\rm{V}_{\rm{NS}}^{\rm{act}}$ is the number of neutrons in the active volume. Rather than using the full volume of a 2 km shell from the surface, we calculate the actual active volume for each neutrino mass based on its specific mean free path. To account for the lower density in the outer layers of a neutron star, we use a mean density of $0.5\rho_0$ along this path. Plugging the numbers and expressions in \eqref{eq:rate}, the total capture rate is found to be
\begin{equation}
    \Gamma\sim10^{25}\,\rm{e}^-/s.
\end{equation}

This corresponds to an energy deposition rate of $\sim 10^{25}\,\rm{MeV}/\rm{s}$, leading to a temperature increase in the outer layers of the neutron star. Continuous cosmic neutrino captures result in a power of $\sim10^{19}\,\rm{erg}/\rm{s}$ due to the persistent emission of 1 MeV electrons, which remain trapped within the neutron star. The actual values, considering the different neutrino masses, are presented in Table \ref{tab:rates}.

\begin{table}[H]
    \centering
    \begin{tabular}{c|c c c}
    \hline
       mass  & $\lambda_{\nu_j}$ [m] & $\Gamma$ [$s^-1$] & P [erg/s]\\
    \hline
       $1\,\rm{meV}$ & $650$  & $7.2\cdot10^{24}$ & $1.4\cdot10^{19}$\\
       $8\,\rm{meV}$ & $250$  & $9.7\cdot10^{23}$ & $1.84\cdot10^{18}$ \\
       $50\,\rm{meV}$ & $550$  & $1.5\cdot10^{23}$ & $2.8\cdot10^{17}$ \\
    \hline
       TOTAL & -- & $8.3\cdot10^{24}$ & $1.6\cdot10^{19}$\\
    \hline
    \end{tabular}
    \caption{Mean free path and capture rates for cosmic neutrinos in neutron-rich medium with a density of $0.5\rho_0$. The last column shows the corresponding power emission from the continuous production of 1 MeV electrons in a neutron star with a 10 km radius.}
    \label{tab:rates}
\end{table}

Using Stefan-Bolztman's law, the achieved total power emission of $1.6\cdot10^{19}\,\rm{erg}/\rm{s}$ results in a temperature increase of $400\, \rm{K}$. For cold neutrons stars with surface temperature of about ${\rm few}\,\, 10^{4}$ K, this corresponds to a 4\% increase. Furthermore, if neutrinos are Majorana particles, the total capture rate and the total power of emission would be $2.3\cdot10^{25}\,\rm{s}^{-1}$ and $4.4\cdot10^{19}\,\rm{erg}/\rm{s}$, respectively; leading to a temperature rise of $500\,$K. This represents a signal enhancement by a factor of 1.3 compared to the Dirac case. If this extra 30\% effect could be measured, it would lead to a diagnosis for Dirac vs. Majorana nature from the sky. 

Depending on the redshift at which neutrinos become non-relativistic, the temperature increase of the NS will scale as $\sim (1+z)$.  It is this slope in the change of temperature of a neutron star that indicates the signature of the heating effect of the C$_{\nu}$B. For the coolest neutron stars, this represents a temperature change from 4\% at $z=0$ to $\sim 40\%$ at $z \sim 10$. Future gravitational wave observatories that measure the temperature of merging neutron stars could detect this effect, even without the need to find an optical counterpart, as the redshift can be inferred from the distance using a cosmology model. 

\subsection{Glancing effect on bare Neutron Stars}

In the case of a naked neutron star~\cite{nakedNS}, meaning one without atmosphere or ocean, our calculations will focus on glancing captures. In this scenario, we will quantify the potential observational signal from neutrino capture events occurring within the last few millimeters of the crust. These events correspond to neutrinos entering the neutron star at angles close to $90^\circ$ with respect to the surface normal, resulting in trajectories close to the surface. This ensures that neutrinos are captured just before leaving the neutron star through the opposite side, in such a way that the forward outgoing electrons produced in the reactions scape the star. 

\begin{figure}[H]
    \centering
    \includegraphics[width=0.65\linewidth]{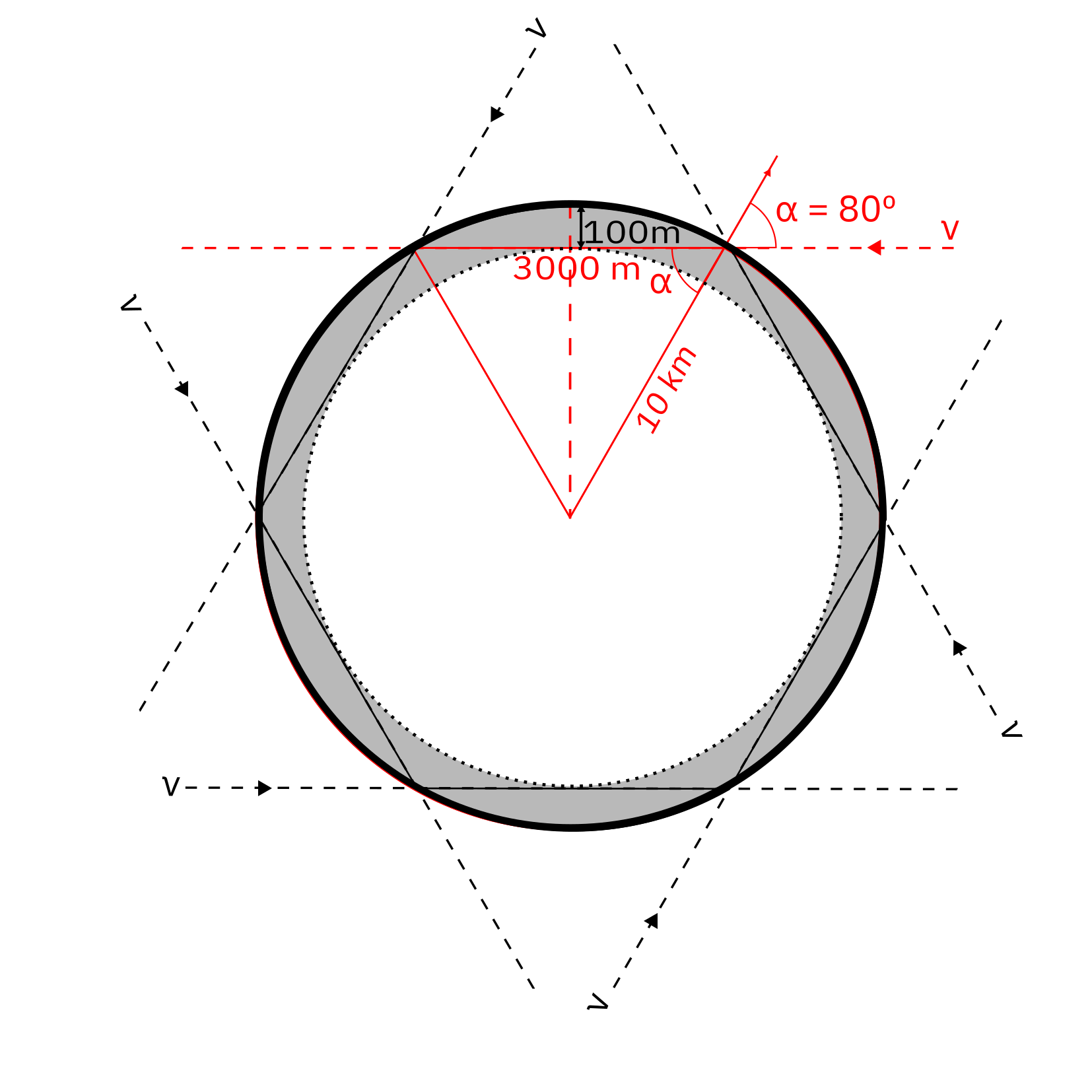}
    \caption{Schematics of the effective volume of the neutron stars that produces the emission of MeV electrons due to C$_{\nu}$B neutrino capture.}
    \label{fig:diagram}
\end{figure}

To avoid absorption, neutrinos cannot reach a certain depth; to remain on the safe side, we assume a maximum depth of 100 m and a mean density of $\rho_{\rm{NS}}=0.1\rho_0$. This corresponds to a neutrino path length of approximately 3 km inside the neutron star, which is on the same order as the mean free path of a cosmic neutrino in a neutron-rich medium with density of $0.1\rho_0$. Only for $m_2$ we set a depth of 50 m and a mean density of $0.05\rho_0$ so the distance inside the star is of the same order of the corresponding mean free path. Calculating this effect implies a neutrino flux reduction which is illustrates in Figure \ref{fig:diagram}. 

We have added a factor $1/2$ to the rate calculation \eqref{eq:rate} to account only for forward outgoing electrons. Detailed numbers of this calculation are displayed in Table \ref{tab:rates2}.

\begin{table}[H]
    \centering
    \begin{tabular}{c|c c c c}
    \hline
       mass  & depth [mm] & $\langle\rho_{\rm{NS}}\rangle$ &  $\lambda_{\nu_j}$ [m] & $\Gamma$ [$s^-1$]\\
    \hline
       $1\,\rm{meV}$ & $100$ & $0.1\rho_0$ & $3200$  & $1.2\cdot10^{23}$ \\
       $8\,\rm{meV}$ & $50$ & $0.05\rho_0$ & $2200$  & $1.0\cdot10^{22}$ \\
       $50\,\rm{meV}$ & $100$ & $0.1\rho_0$ & $2800$  & $2.8\cdot10^{21}$ \\
    \hline
       TOTAL & -- & -- & -- & $1.3\cdot10^{23}$ \\
    \hline
    \end{tabular}
    \caption{Capture rates for glancing cosmic neutrinos in a neutron-rich medium, computed such that the penetration depth corresponds to a path length through the neutron star comparable to the mean free path of each neutrino mass eigenstate in a neutron-rich medium of a given density that varies with depth, affecting the mean free path accordingly.}
    \label{tab:rates2}
\end{table}

Using the values from Table \ref{tab:rates2}, the total capture rate due to this effect is found to be $\Gamma=10^{23}\,e/s$. As before, the outgoing electrons produced by the cosmic neutrino capture reaction have a kinetic energy of 1 MeV, meaning that the luminosity of the expected line is $10^{17}$ egs s$^{-1}$.

\section{Planets}
\label{sec:planets}

The second type of astrophysical objects proposed are planets through which neutrinos can pass without interaction, meaning the flux is not reduced. Additionally, there are planets without atmospheres, allowing electrons produced on the surface via neutrino capture to escape, making an eventual signal detectable. We will calculate the capture on potassium-40, one of the most abundant and long-lived beta decaying nuclei. After examining the single-beta decaying isotopes of the most abundant elements on Earth, we concluded that the dominant factor for neutrino capture is abundance rather than cross-section. Although isotopes such as $^{32}$Si, $^{45}$Ca, or $^{24}$Na have larger cross-sections, their shorter lifetimes mean they are only present in trace amounts, as abundance decreases exponentially with mean life. For this reason, potassium-40 is the dominant source. The capture reaction is
\begin{equation}
    \nu_j + ^{40}\rm{K} \rightarrow ^{40}\rm{Ca} + e^- .
\end{equation}
For this particular case, the capture is much more suppressed than the capture of a neutron, because the beta decay of potassium is a third-forbidden reaction. The cross-section can be calculated from the half-life of the beta decay, as proposed in \cite{A.G.Cocco_2008}:

\begin{equation}
    \sigma\, v_\nu = 2\pi^2\ln2\frac{p_eE_eF(Z,E_e)}{ft_{1/2}}.
    \label{eq:sec}
\end{equation}
For $^{40}\rm{K}$, the cross-section times velocity is $8\cdot10^{-57}\,\rm{cm}^2$, 13 orders of magnitude smaller than the capture on neutrons. We can use our Earth as a test case; potassium is one of the most abundant elements on the Earth's surface, particularly in water. The abundance in water is $400 \,\rm{mg}/\rm{L}$, plus the natural abundance of the $^{40}\rm{K}$ isotope $(0.012\%)$, resulting in a total abundance of $\sim 10^{15}\,\rm{nuclei}/\rm{cm}^3$. 

Equally to the case of neutron stars, the outgoing electron has a kinetic energy of $\sim 1\,\rm{MeV}$, so for an electron signal, the active volume is limited to just the outermost centimeters of the planet. If we take Earth as an example with a thickness $\Delta R = 1 \,\rm{cm}$, the active volume would be
\begin{equation}
    V^{act}_\oplus=4\pi R_\oplus^2\Delta R \sim 5\cdot10^{18}\,\rm{cm^3}.
\end{equation}
Hence:
\begin{equation}
    \Gamma = 8\cdot10^{-57}\rm{cm}^2\cdot\, 3\cdot10^{10}\frac{cm}{s} \cdot 56\frac{\nu}{cm^3}\cdot 10^{15} \frac{nuclei}{cm^3} \cdot 5\cdot10^{18} \rm{cm}^3 = 0.002\, \rm{e}/\rm{yr}.
\end{equation}

This would represent the total capture in a liquid medium with dissolved potassium that is permeable to $1 - 2 \,\rm{cm}$ to $1\,\rm{MeV}$ electrons. In the case of hypothetical exo-water worlds with radii much larger than Earth, this could increase the signal to close to one electron per year. The signal is feeble and most likely not detectable with currently envisioned technology; future will tell if this could be measure\footnote{``I have done a terrible thing: I have postulated a particle that cannot be detected." Wolgnag Pauli (1930).}.

\section{Conclusion}

We have computed in detail the effect that the capture of C$_{\nu}$B would produce in astrophysical objects, with novel observable effects. The most obvious candidates are those that contain single beta decay species. These are neutron stars and planets with heavy element composition. In the case of neutron stars, the signal of an increase of the surface temperature of order \% or in the case of naked neutron stars in a line of emission in the MeV range. An obvious observational strategy for discovery would be the stacking of the spectrum of multiple neutron stars in the Milky Way, which contains nearly $100$  million neutron stars. A novel signature is the redshift dependence of the temperature heating, which could be detected from future gravitational waves observatories \cite{ET:2019dnz} that follow neutron star mergers. For the case of planets, the signal is feeble and very challenging. Some increase to the signal could come from much larger reservoirs of potassium than Earth in exoplanets and a stacking of the signal from these. Our calculations show that the direct detection of the C$_{\nu}$B remains challenging, but emphasis on the signal in neutron stars could yield challenging but realistic chances of detection.

\bibliographystyle{JHEP}

\providecommand{\href}[2]{#2}\begingroup\raggedright\endgroup

\end{document}